\documentclass[numberedappendix]{emulateapj}

\newcommand{\himpc}{{\hbox {$~h^{-1}$}{\rm ~Mpc}}}

\slugcomment{Accepted to \textit{The Astrophysical Journal} 12/18/2008}

\shorttitle{Misalignment of LRGs with their Host Dark Matter Halos}
\shortauthors{Okumura, Jing, \& Li}

\begin{document}

\title{Intrinsic Ellipticity Correlation of SDSS Luminous Red Galaxies \\ 
and Misalignment with their Host Dark Matter Halos}

\author{Teppei Okumura\altaffilmark{1}, Y. P. Jing\altaffilmark{1},
and Cheng Li\altaffilmark{1,2}}

\email{teppei@shao.ac.cn} 

\altaffiltext{1} {Key Laboratory for Research in Galaxies and
Cosmology, Shanghai Astronomical Observatory, Chinese Academy of
Sciences, 80 Nandan RD, Shanghai¡¤200030¡¤China}

\altaffiltext{2} {Max-Planck-Institut f\"ur Astrophysik,
Karl-Schwarzschild-Strasse 1, 85748 Garching, Germany}

\begin{abstract}
We investigate the orientation correlation of giant elliptical
galaxies by measuring the intrinsic ellipticity correlation function
of 83,773 luminous red galaxies (LRGs) at redshifts 0.16 -- 0.47 from
the Sloan Digital Sky Survey.  We have accurately determined the
correlation up to 30 $\himpc$. Luminosity dependence of the
ellipticity correlation is also detected although the error bars are
large, while no evidence is found for its redshift evolution between
$z=0.2$ and $z=0.4$.  Then we use a cosmological $N$-body simulation
to examine misalignment between the central LRGs and their parent dark
matter halos.  Central and satellite galaxies are assigned to
simulated halos by employing a halo occupation distribution model for
the LRGs. The ellipticity correlation is predicted to have the same
shape as but an amplitude about 4 times higher than our observation if
the central LRGs are perfectly aligned with their host halos. This
indicates that the central LRG galaxies are preferentially but not
perfectly aligned with their host halos. With the assumption that
there is a misalignment angle between a central LRG and its host halo
which follows a Gaussian distribution with a zero mean and a width
$\sigma_\theta$, we obtain a tight constraint on the misalignment
parameter, $\sigma_\theta={35.4}^{+4.0}_{-3.3}$ deg. This type of
intrinsic ellipticity correlation, if not corrected, can lead to
contamination at 5\% level to the shear power spectrum in weak
lensing surveys of limiting magnitude $R_{AB}=24.5$ if the source
central galaxies follow the same misalignment distribution as the
LRGs. 
\end{abstract}


\section{Introduction}
Weak gravitational lensing by large-scale structure provides a unique
tool that directly probes matter distribution in the universe.  One of
the most serious contaminations for weak lensing observations comes
from two types of intrinsic alignments: the ellipticity correlation of
source galaxies with each other (intrinsic ellipticity--intrinsic
ellipticity correlation) and the ellipticity correlation of source
galaxies with the surrounding matter distribution (gravitational
shear--intrinsic ellipticity correlation).

There was much work based on both analytical and numerical methods
which attempted to estimate the intrinsic ellipticity--intrinsic
ellipticity correlation \citep{Heavens2000, Croft2000, Lee2000,
Catelan2001, Crittenden2001, Lee2001, Jing2002}.  According to these
previous studies the effect of intrinsic alignment can lead to $\sim
10 \% $ or even higher contamination for a deep survey with median
source redshift of 1 if galaxies are aligned with the angular momentum
or the ellipticity of their host halos. Fortunately, this effect can
be reduced by downweighting nearby source pairs with either
spectroscopic or photometric redshifts \citep{King2002, Heymans2003,
Takada2004,King2005,Fan2007}.  Although the intrinsic
ellipticity--intrinsic ellipticity correlation has been detected in
several observations at low redshift \citep{Pen2000, Bernstein2002,
Brown2002, Hirata2004b, Heymans2004, Lee2007}, the amplitude of the
correlation is much smaller than theoretical predictions
\citep{Heymans2004, Mandelbaum2006b, Heymans2006}, which indicates
that galaxies are not perfectly aligned with the angular momentum or
the ellipticity of their host halos. \citet{Heymans2004} explained the
discrepancy in amplitude of the ellipticity correlation function
between the model predictions and the observations by assigning a
random misalignment angle around the original halo angular momentum
vector.  On the other hand, \citet{Hoekstra2004} detected the
significant flattening of dark matter halos along the minor axes of
galaxies from weak lensing analysis, which implies that the halos are
well aligned with the major axes of the galaxies. In a later work,
however, \citet{Mandelbaum2006a} did not detect this effect even using
a much larger SDSS data set.

There are also observational studies which have detected with a high
confidence the gravitational shear--intrinsic ellipticity correlation
\citep{Mandelbaum2006b, Hirata2007}, although downweighting this
effect is more complicated \citep{Hui2008, Hirata2004a, Heymans2006,
Bridle2007, Joachimi2008}. In this paper, we focus only on the
intrinsic ellipticity--intrinsic ellipticity correlation, so we call
it the intrinsic ellipticity correlation for brevity.

Meanwhile, investigating intrinsic alignment of galaxies relative to
their host halos is also important because it contains information
about galaxy formation and evolution. Recently there has been
increasing interest in the misalignment between central galaxies and
their parent dark matter halos. It was shown in previous studies based
on $N$-body simulations that the angular momentum distributions of gas
and dark matter components are partially aligned, with a typical
misalignment angle of $\sim 30^{\circ}$\citep{van den Bosch2002,
Chen2003, Sharma2005}, but this might be relevant to disk galaxies
only. It is also expected that the central ellipticals are aligned
with their host halos to a certain degree if the ellipticals are
formed by dry mergers \citep{Dubinski1998,Naab2006,
Boylan-Kolchin2006}, because the orientations of the central
ellipticals and of the host dark matter halos are respectively
determined by the orbital angular momenta of their progenitor galaxies
and halos that are correlated. In observation, by studying the
alignment of central galaxies with their satellite spatial
distributions in SDSS groups \citep{Yang2006} and by assuming that the
satellites follow dark matter in spatial distribution,
\citet{Kang2007} and \citet{Wang2008} have reached somewhat
conflicting conclusions about the misalignment angle between the
central galaxies and their host halos (typically $40^{\circ}$ in
\citet{Kang2007} and $23^{\circ}$ in \citet{Wang2008}).

In this paper we present the ellipticity correlation functions of a
spectroscopic luminous red galaxy (LRG) sample from the Sloan Digital
Sky Survey \citep[SDSS;][]{Y2000}. We estimate the luminosity and
redshift dependences of the ellipticity correlations. LRGs are
massive, and most of them are located in the central regions of rich
groups or galaxy clusters. A small fraction of satellites can be
reliably identified in the observation. Therefore, we are able to
study the misalignment between central LRGs and their parent dark
halos by comparing the observed ellipticity correlation function with
that of the dark halos in an $N$-body simulation.

Compared to previous work concerning the misalignment, our analysis
using the LRG sample have at least three advantages.  First, the
orientation of satellite galaxies relative to their host halos might
be very different from that of the central galaxies. Central LRGs are
easy to be identified, which enables us to reliably determine the
misalignment of central galaxies with their host halos without
contamination from satellite galaxies. Second, all LRG galaxies are
believed to be the product of dry mergers, and their formation
processes are distinct from those of spiral (disk) galaxies. Our
analysis will be on central ellipticals without contamination from
disk galaxies. Finally, because LRGs preferentially reside in massive
halos and such host halos have stronger ellipticity correlations than
less massive ones hosting fainter galaxies \citep{Jing2002}, it is
easier to accurately measure their ellipticity correlations and
determine their misalignment angle relative to their host halos.

The structure of this paper is as follows.  In Section \ref{sec:sdss},
we describe the SDSS LRG sample used in our analyses. We measure the
ellipticity correlation functions of SDSS LRGs in Section
\ref{sec:ellip}.  The luminosity and redshift dependences of the
ellipticity correlation function are also presented.  In Section
\ref{sec:model} we calculate the model ellipticity correlation
functions of LRGs using an $N$-body simulation with the assumption
that central LRGs are completely aligned with their parent dark matter
halos.  We examine and constrain the misalignment between central LRGs
and their parent halos in Section \ref{sec:misalignment}.  Our
conclusions are given in Section \ref{sec:conclusion}.

\section{SDSS Luminous Red Galaxy Sample} \label{sec:sdss}

We analyze the LRG sample from the SDSS \citep{Y2000, Stoughton2002}.
The LRG selection algorithm \citep{E2001} selects $\sim 12$ galaxies
per square degree using color and magnitude cuts.  The resulting
galaxies have a Petrosian magnitude $r<19.5$, which tend to be
luminous early-types and to be located in rich groups or clusters of
galaxies. All fluxes are corrected for reddening \citep{Schlegel1998}
before use.  The LRG selection is so efficient that it produces a
volume-limited sample, and thus the comoving number density of the
sample is close to a constant out to $z\sim 0.36$ and drops thereafter
due to the flux limits \citep[see Figure 1 of][]{Zehavi2005}.

For our analysis we use 83,773 LRGs in the redshift range from 0.16 to
0.47 from the SDSS Data Release 6 \citep{Adelman-McCarthy2008} which
is publicly available.  We choose only the LRGs for which the redshift
confidence parameter is greater than 0.95.  The galaxies in the sample
have rest-frame {\it g}-band absolute magnitudes $-23.2<M_{g}<-21.2$
($H_0=100\mbox{~km~s}^{-1}\mbox{~Mpc}^{-1}$) with $K + E$ corrections
of passively evolved galaxies to a fiducial redshift of $0.3$
\citep[see Appendix B of][]{E2001}.

The goal of this study is to investigate the misalignment of central
galaxies and their host dark matter halos.  Accurate measurement of
redshifts enables us to divide the LRGs in our sample into centrals
and satellites. Following \citet{Reid2008}, we adopt criteria of
$r_{\perp}\leq 0.8 \himpc$ and $r_{\parallel}\leq 20\himpc$ for two
galaxies to be in the same halo, where $r_{\perp}$ and $r_{\parallel}$
are their separations perpendicular and parallel to the line of sight,
respectively. The criteria imply that 79,493 LRGs (94.9\%) are
centrals, which is consistent with the results not only from mock LRGs
described in Section \ref{sec:hod} but also from observational work by
\citet{Zheng2008} and \citet{Reid2008}. For the halos with two or more
LRGs, we regard the brightest one as the central and the others as the
satellites.

Besides a sample of LRGs, we also need information on their
shapes. There are several model-dependent and model-independent ways
in the SDSS data to measure the ellipticity of galaxies
\citep{Stoughton2002}. Following the previous work which investigated
the central-satellite alignment \citep[e.g.,][]{Brainerd2005,
Yang2006, Faltenbacher2007}, we adopt the latter and define the
ellipticity of galaxies with the ellipticity of the 25 mag
arcsec${}^{-2}$ isophote in the $r$ band.  In addition, the
point-spread function (PSF) has been corrected when measuring galaxy
shapes in the SDSS imaging pipelines \citep{Fischer2000, Lupton2001,
Stoughton2002}.  More accurate schemes for correcting the PSF were
adopted in previous work of weak lensing \citep{Hirata2004b,
Mandelbaum2005}. However, the original correction for the PSF in the
SDSS should be sufficient for the current analysis because only the
position angles of LRGs are used in most of our study.

\section{Ellipticity Correlation of LRGs}\label{sec:ellip}

\subsection{Measuring the LRG Ellipticity Correlation Functions}
\label{sec:ecf}
The ellipticity correlation is examined by measuring the ellipticity
of each galaxy.  The two components of the ellipticity are defined as
\begin{eqnarray}
  \left(
  \begin{array}{c} e_1 \\ e_2 \end{array}
  \right) = \frac{1-q^2}{1+q^2} \left(
  \begin{array}{c}\cos{2\beta} \\ \sin{2\beta}\end{array}
  \right), \label{eq:ellip}
\end{eqnarray}
where $q$ is the ratio of minor and major axes ($0\leq q \leq 1$) and
$\beta$ is the position angle of the ellipticity from the north
celestial pole to east \citep{Stoughton2002}. Then the ellipticity
correlation functions are defined \citep[e.g.,][]{Miralda-Escude1991,
Heavens2000, Croft2000, Jing2002} as
\begin{equation}
  c_{ab}(r)=\left\langle e_a({\bf x})e_b({\bf x}+{\bf
    r})\right\rangle, \label{eq:ecf}
\end{equation}
where ${\bf r}$ is the three-dimensional vector of separation $r$
joining a pair of galaxies. The comoving distance to every galaxy,
$x(z)$, is calculated by assuming a flat universe with $\Omega_m =
1-\Omega_\Lambda = 0.3$, where $\Omega_m$ and $\Omega_\Lambda$ are the
mass density parameter and the cosmological constant parameter,
respectively. In Eq.(\ref{eq:ecf}) the components of the ellipticity,
$e_1$ and $e_2$, are redefined by rotating by an angle between the
north pole and the line connecting the two galaxies on the celestial
sphere. Thus, the ellipticity component $e_1$ ($e_2$) corresponds to
the elongation and compression along (at $45^{\circ}$ from) the line
joining the two galaxies.

\begin{figure}[tb]
\epsscale{1.1}
\plotone{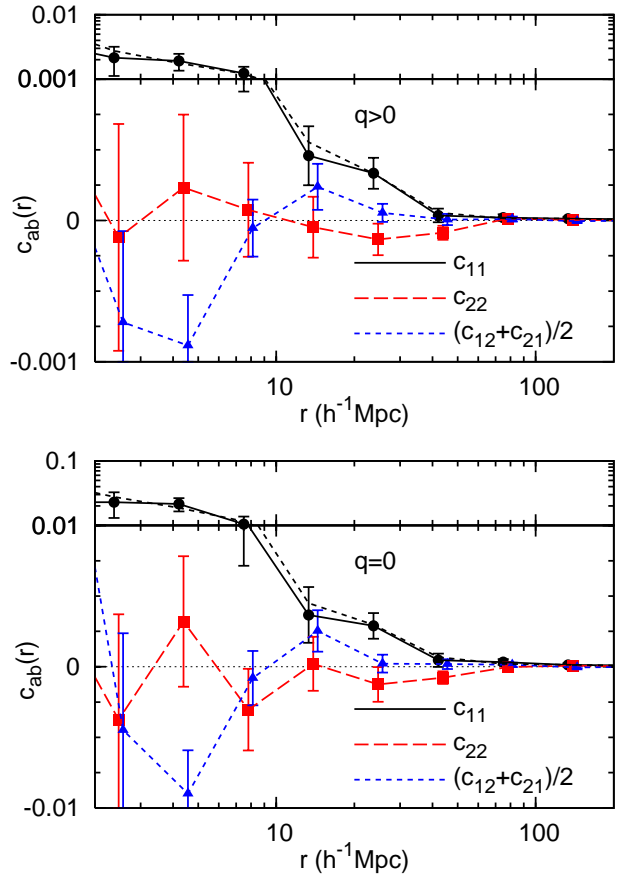}
\caption{Top: ellipticity correlation functions for the central LRG
sample. Bottom: as the top panel, but setting $q=0$ when the
ellipticity of the LRGs is measured in equation (\ref{eq:ellip}).  The
dashed black line shows $c_{11}$ for the combined sample of central
and satellite LRGs. To clearly show the fluctuations of $c_{22}$ and
$(c_{12}+c_{21})/2$, mixed logarithmic and linear scalings are used
for the vertical axis. Bins in $r$ are in logarithmic separation of
0.25. The circles/triangles have been respectively offset in the
negative/positive direction for clarity. \\~}
\label{fig:cij}
\end{figure}

The resulting ellipticity correlation functions for the observed
central LRGs (see Section \ref{sec:sdss}) are shown in the top panel of
Figure \ref{fig:cij}.  For comparison, the result of $c_{11}$ measured
from all the LRGs including satellites is also given, which shows that
the difference between the results with and without satellites is very
small, because the contribution from the satellites is negligible. The
error bars shown in the figure represent $1\sigma$ errors estimated
with the jackknife resampling method. A description of the method is
given at the end of this section.

%%%%%%%%%%%%%%%%%%%%%%%%%%%%%%%%%%%%%%%%%%%%%%%%%%%%%%%
%%%%%%%%%%%%%%%%%%%%%%%%%%%%%%%%%%%%%%%%%%%%%%%%%%%%%%%

As for the auto-correlation functions, while we clearly detect the
positive correlation of ellipticity in $c_{11}$ particularly on scales
less than $30\himpc$, the amplitude of $c_{22}$ is found to be much
smaller. It is simply because $c_{11}$ is nearly isotropic but
$c_{22}$ is very anisotropic \citep{Croft2000, Jing2002}; thus the
amplitude of $c_{22}$ is suppressed when averaged over different
directions.  On the other hand, the cross-correlations $c_{12}$ and
$c_{21}$ should vanish on all scales \citep{Heavens2000, Croft2000,
Jing2002} and our results observationally confirm that
$(c_{12}+c_{21})/2$ fluctuates around zero within the measurement
errors except for a point at $r\approx 4\himpc$ with a $2\sigma$
deviation .  In the following analysis the function $c_{11}$ is
mainly discussed and $c_{22}$ is used only for a cross-check of the
results obtained from the measurement of $c_{11}$.

The bottom panel of Figure \ref{fig:cij} is the same as the top panel,
except that it shows the results when the axis ratio $q$ in equation
(\ref{eq:ellip}) is set to be zero. It is equivalent to assuming that
a galaxy is a line along its major axis, and the measurement indicates
the correlation between the orientations of two galaxies with their
shape not being considered. This prescription is important because
only information on position angles is necessary when we examine
the misalignment between LRGs and their parent dark halos.  Throughout
this paper, we perform all the statistical analyses using the
ellipticity correlation functions with $q=0$. We note that this
$c_{11}$ with $q=0$ is about 10 times larger than that with $q \ne 0$,
consistent with the fact that the median value of $q$ is 0.73 for the
sample.

Statistical errors on the measurement of the ellipticity correlation
functions are estimated using jackknife resampling
\citep[e.g.,][]{Lupton1993} with 99 angular subsamples.  Because the
number of data points used for our statistical analysis is 8, this
number of subsamples is large enough to obtain a nonsingular matrix.
Each subsample includes a region contiguous on the sky, the comoving
size of which is about $120\himpc$ on a side at $z=0.3$.

\begin{figure}[tb]
\epsscale{1.1}
\plotone{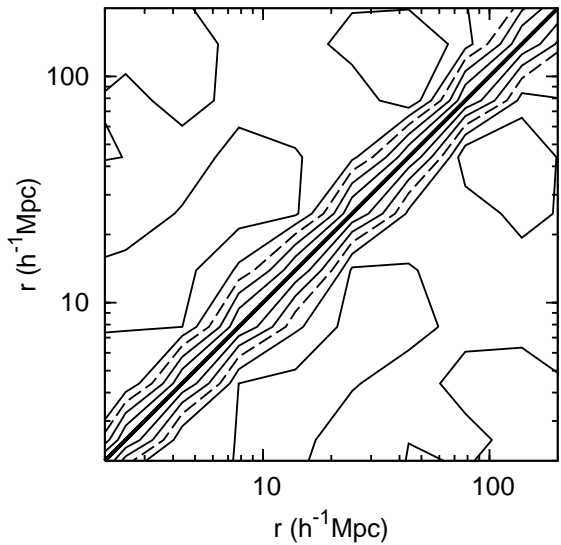}
\caption{Normalized covariance matrix for the $c_{11}(r)$ measurement
of the central LRG sample. Contour spacing is 0.2 going from 1 on the
diagonal (thick line) down to 0. The dashed line denotes the 0.4
contour. This is a similar plot to Figure 7 of \citet{Zehavi2005} who
focused on the LRG clustering.\\~}
\label{fig:cov}
\end{figure}

We obtain the covariance matrix of $c_{11}$ for the central LRGs from
the jackknifed realizations by
\begin{equation}
  C_{ij} =
\frac{N-1}{N}\sum^N_{l=1}\left(c_{11}^l(r_i)-\bar{c}_{11}(r_i)\right)
\left(c_{11}^l(r_j)-\bar{c}_{11}(r_j)\right),
  \label{eq:cov}
\end{equation}
where $N=99$, $c^l_{11}(r_i)$ represents the value of $c_{11}(r)$ of the
$i$th separation bin in the $l$th realization, and $\bar{c}_{11}(r_i)$ is
the mean value of $c_{11}(r_i)$ over all realizations. Figure
\ref{fig:cov} shows the obtained covariance matrix normalized by the
diagonal elements, $C_{ij}/(C_{ii}\cdot C_{jj})^{1/2}$. As is clearly
seen, almost all the contribution of statistical errors comes from the
diagonal elements. The error bars shown in Figure \ref{fig:cij} are a
square root of the diagonals of the matrix, $C_{ii}^{1/2}$.  We have
tried several values of $N$, and found that the error bars we
obtain are very stable against the changes of $N$.

%%%%%%%%%%%%%%%%%%%%%%%%%%%%%%%%%%%%%%%%%%%%%%%%%%%%%%%
%%%%%%%%%%%%%%%%%%%%%%%%%%%%%%%%%%%%%%%%%%%%%%%%%%%%%%%

\subsection{Luminosity and Redshift Dependences}
\label{sec:dependence}

Before proceeding to the next section, we examine the dependences of
the measured ellipticity correlation function of LRGs on their
luminosity and redshift. Previous studies found a luminosity
dependence for the clustering of LRGs \citep{Zehavi2005,
Percival2007}, which indicates that LRGs of different luminosities are
in halos of different mass. Because the ellipticity correlation of
halos increases with the halo mass \citep{Jing2002}, it is naturally
expected that there also exist the luminosity dependence for the
ellipticity correlations of galaxies.

\begin{figure}[tb]
\epsscale{1.1}
\plotone{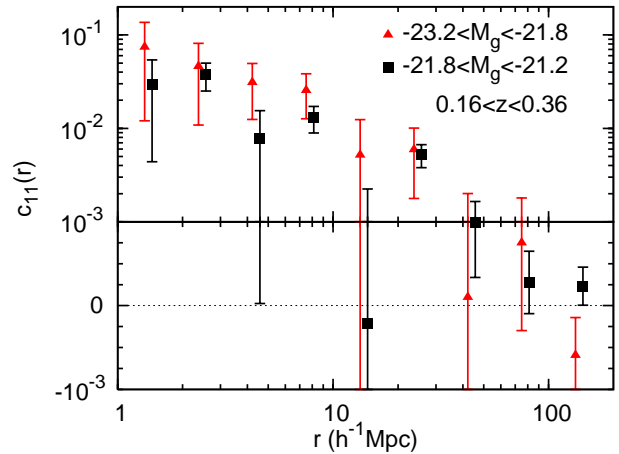}
\caption{Luminosity dependence of $c_{11}(r)$ of the LRGs for
$0.16<z<0.36$. Here both the central and satellite LRGs are used for
the calculation. Note that the vertical axis mixes logarithmic and
linear scalings. \\~}
\label{fig:ecf_luminosity}
\end{figure}

Figure \ref{fig:ecf_luminosity} shows $c_{11}(r)$ for two spans of
$M_g$ at $0.16<z<0.36$. The magnitude cuts are similar to those
adopted in \citet{Zehavi2005} and the redshift range is chosen to
utilize a volume-limited sample. Stronger correlation of ellipticity
can be seen in the brighter sample ($-23.2<M_g<-21.8$) on small scales
although the error bars are large because of the sparseness and
limited survey volume of the sample. This result is consistent with
the expectation that the more luminous LRGs are located in more
massive halos \citep{Zehavi2005, Zheng2008} that have a stronger
ellipticity correlation function \citep{Jing2002}.

%%%%%%%%%%%%%%%%%%%%%%%%%%%%%%%%%%%%%%%%%%%%%%%%%%%%%%%
%%%%%%%%%%%%%%%%%%%%%%%%%%%%%%%%%%%%%%%%%%%%%%%%%%%%%%%

\begin{figure}[tb]
\epsscale{1.1}
\plotone{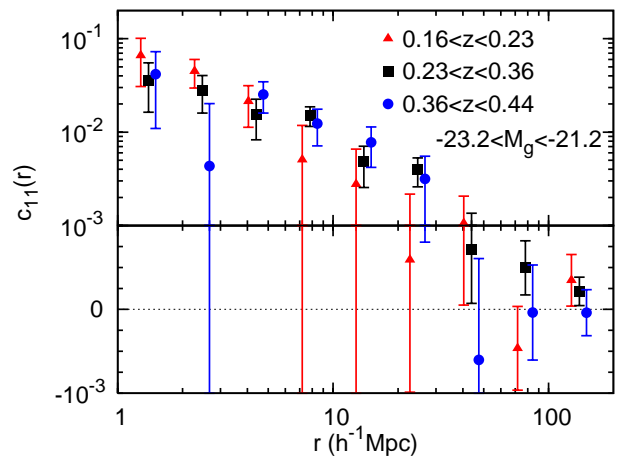}
\caption{Redshift dependence of $c_{11}(r)$ of the LRGs for the
absolute magnitude of $-23.2<M_g<-21.2$. Here both the central and
satellite LRGs are used for the calculation. Note that the vertical
axis mixes logarithmic and linear scalings. \\~}
\label{fig:ecf_redshift}
\end{figure}

\citet{Lee2008} analyzed simulation data in order to pursue the
nonlinear evolution of galaxy intrinsic alignment at $0<z<2$. They did
not find an indication for the redshift evolution of the ellipticity
correlation although they could detect the evolution of its
nonlinearity.  Thus it may be difficult for current observations to
detect the redshift dependence of the ellipticity correlation.
Redshift evolution for the LRG clustering was also not detected
\citep{Zehavi2005}.  Figure \ref{fig:ecf_redshift} shows the redshift
dependence of $c_{11}(r)$.  Because the redshift range of our LRG
sample is not large, even shallower than that examined in
\citet{Lee2008}, we do not find such a dependence in the LRG sample.
 
%%%%%%%%%%%%%%%%%%%%%%%%%%%%%%%%%%%%%%%%%%%%%%%%%%%%%%%
%%%%%%%%%%%%%%%%%%%%%%%%%%%%%%%%%%%%%%%%%%%%%%%%%%%%%%%

\section{Model Predictions}\label{sec:model}

\subsection{$N$-body Simulation}\label{sec:nbody}
To make model predictions for the ellipticity correlation function, we use
a halo catalog constructed from a high-resolution cosmological
simulation with $1024^3$ particles in a cubic box of side $1200\himpc$
\citep{Jing2007}.  A spatially flat $\Lambda$CDM model with
$\Omega_m=1-\Omega_\Lambda=0.268$, $\Omega_b=0.045$, $h=0.71$, and
$\sigma_8=0.85$, was assumed, where $\Omega_b$ is the baryon density
parameter, $h$ is the Hubble parameter normalized by
100~km~s${}^{-1}$Mpc${}^{-1}$, and $\sigma_8$ is the present linear
rms density fluctuation within a sphere of radius $8\himpc$.  Dark
matter halos are identified in the $z=0.274$ output using the
friends-of-friends algorithm with a linking length $b$ equal to $0.2$
times the mean particle separation.  See
\citet{Jing2007} for details of the simulation.

\subsection{Halo Occupation Distribution}\label{sec:hod}
In order to assign galaxies to the simulated halos, we rely on the
framework of the halo occupation distribution \citep[HOD,
e.g.,][]{Jing1998, Ma2000, Peacock2000, Seljak2000, Scoccimarro2001,
Berlind2002}, which describes the relationship between the galaxy and
dark matter density fields.  HOD modeling has been performed for LRG
galaxies by several independent approaches \citep{Masjedi2006, Ho2007,
Blake2008, Kulkarni2007, White2007, Seo2008, Zheng2008, Reid2008}.

A flexible parameterization with five HOD parameters was introduced by
\citet{Zheng2005} \citep[see also][]{Zheng2007}. The mean occupation
function of galaxies within a dark halo of mass $M$, being the sum of
central and satellite mean occupation functions, is parameterized as
\begin{eqnarray}
  &&\left\langle N(M) \right\rangle = \left\langle N_{\rm cen}(M)
  \right\rangle \left(1+\left\langle N_{\rm sat}(M) \right\rangle
  \right), \\ &&\left\langle N_{\rm cen}(M) \right\rangle =
  \frac{1}{2}\left[ 1+ {\rm erf}\left(\frac{\log{M}-\log{M_{\rm min}}}
  {\sigma_M}\right) \right], \nonumber \\ &&\left\langle N_{\rm
  sat}(M) \right\rangle = \left(\frac{M-M_0}{M_1'}\right)^{\alpha},
  \nonumber
\end{eqnarray}
where erf is the error function, $M_{\rm min}$ is the characteristic
minimum mass to host a central galaxy, $M_1'$ is a mass for a halo
with a central galaxy to host one satellite when $M_0 \ll M_1'$, $M_0$
is the mass scale to truncate satellites, and $\sigma_M$ is the
characteristic transition width.  Following the latest fits for the
HOD parameters of the LRGs by \citet{Seo2008}, we choose $M_{\rm
min}=8.226\times 10^{13}M_{\sun}$, $M_1'=6.875\times 10^{14}M_{\sun}$,
$M_0=3.209\times 10^{9}M_{\sun}$, $\sigma_M=0.556$, and $\alpha=1.86$.
A central LRG is assigned to a halo based on the nearest integer
distribution with the average of $\langle N_{\rm
cen}(M)\rangle$. Satellite LRGs are then assigned to each halo with a
central based on the Poisson distribution with the average of $\langle
N_{\rm sat}(M)\rangle$.  The satellite LRGs inside dark matter halos
are distributed following the Navarro-Frenk-White profile
\citep{Navarro1997}. The
resulting fraction of central LRGs is 93.7\%, consistent with that
from the observation (Section \ref{sec:sdss}).

\begin{figure}[tb]
\epsscale{1.1}
\plotone{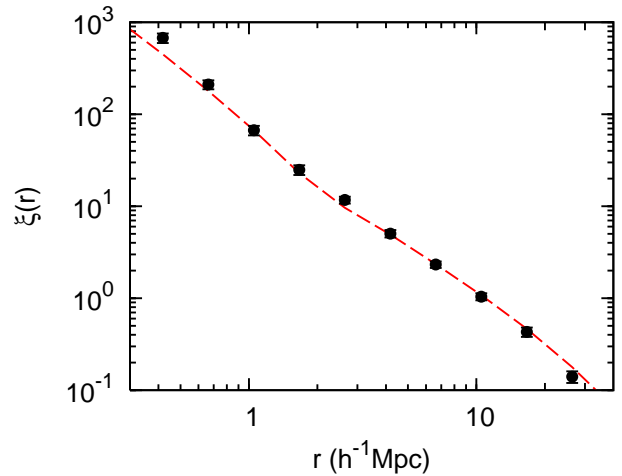}
\caption{Comparison of the real space correlation functions between
the observed and mock LRGs. The black points with the error bars show
the observed correlation function \citep{Zehavi2005}. The dashed line
is that of the mock galaxy catalog using the best-fit HOD model for
the LRGs \citep{Seo2008}. \\~}
\label{fig:xi}
\end{figure}

In Figure \ref{fig:xi}, we show a comparison of the real-space
correlation function between the mock and observed \citep{Zehavi2005}
LRGs. Very good agreement of the results between the observation and
mock catalog can be seen except for $r<0.5 \himpc$, as was seen by
\citet{Seo2008}.  This small discrepancy is irrelevant to the
current study because the satellite distribution within halos
dominates on this scale and only central LRGs are used for the
statistical analysis below.
%%%%%%%%%%%%%%%%%%%%%%%%%%%%%%%%%%%%%%%%%%%%%%%%%%%%%%%
%%%%%%%%%%%%%%%%%%%%%%%%%%%%%%%%%%%%%%%%%%%%%%%%%%%%%%%

\subsection{Modeled Ellipticity Correlation Function}\label{sec:hecf}

The principal axes of each halo in a projected plane are computed by
diagonalizing the momentum of inertial tensor
\citep[e.g.,][]{Miralda-Escude1991, Croft2000}
\begin{equation}
  I_{ij}=\sum x_i x_j,
\end{equation}
where the sum is over all the particles in the halo. The ellipticity
components of each halo are then estimated in the same way as those of
LRGs (eq. [\ref{eq:ellip}]), where the value of $q$ is assumed to be
zero again.

\begin{figure}[tb]
\epsscale{1.1}
\plotone{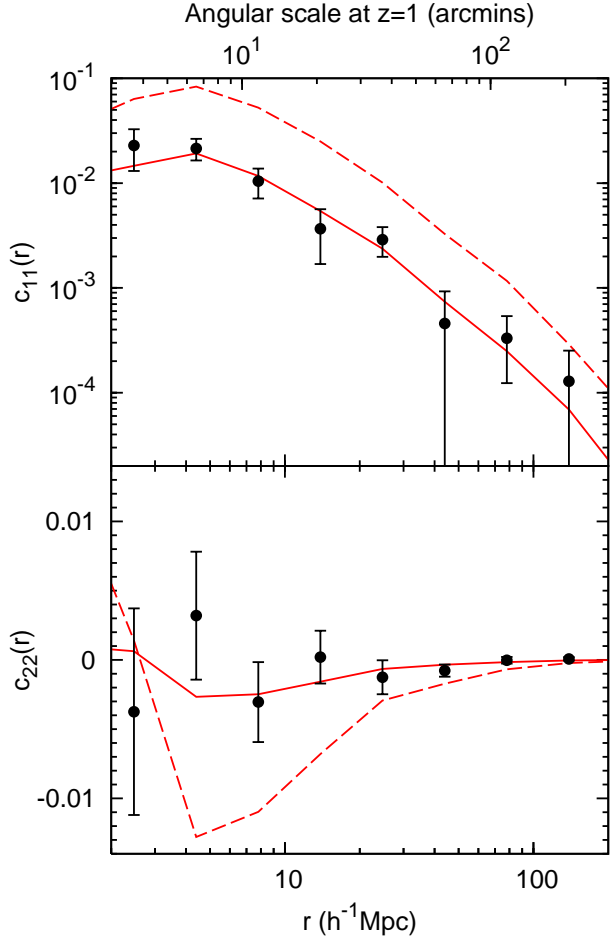}
\caption{Ellipticity auto-correlation functions of the central LRGs,
(top) $c_{11}(r)$ and (bottom) $c_{22}(r)$. In both panels, the data
points with the error bars are the measurements from the SDSS, the
same ones as those in the bottom panel of Figure \ref{fig:cij}. The
dashed red lines are results of the mock central LRGs with no
misalignment with their parent halos. The solid red lines are those
with the misalignment parameter of $\sigma_\theta = 35^{\circ}$.  The
horizontal axis at the top shows the corresponding angular scale when
all the galaxies are located at $z=1$.  \\~}
\label{fig:vardelta}
\end{figure}

First we assume that all central galaxies are completely aligned with
their parent dark matter halos. Then the ellipticity correlation
functions of central galaxies are equal to those of their parent
halos. With this assumption, we plot the ellipticity auto-correlation
functions of the mock LRGs, $c_{11}$ and $c_{22}$, in Figure
\ref{fig:vardelta}.  In order to refine the statistics, we averaged
over seven mock LRG samples with different random seeds for assigning
LRGs to dark halos.  Interestingly, the ellipticity correlation
function $c_{11}$ of the mock LRGs has a very similar shape to the
observed function, but the amplitude is about 4 times higher. The
function $c_{22}$ is significantly negative at $r$ about a few
$\himpc$, compared to the real observed one. In the next section we
will explain these differences between the observation and simulation
by considering misalignment of central galaxies with their host halos.
In Figure \ref{fig:vardelta} we also show the angular separations with
the assumption of all the galaxies being at $z=1$, which is the
typical redshift of recent weak lensing surveys (see Section
\ref{sec:conclusion}).  Note that the values of the ellipticity
correlation function of halos are about an order of magnitude larger
than the previous result by \citet{Jing2002}, because we assume $q=0$
in the current study.

%%%%%%%%%%%%%%%%%%%%%%%%%%%%%%%%%%%%%%%%%%%%%%%%%%%%%%%
%%%%%%%%%%%%%%%%%%%%%%%%%%%%%%%%%%%%%%%%%%%%%%%%%%%%%%%

\section{Constraints on Misalignment}\label{sec:misalignment}
In this section we consider a more general case in which the position
angle of each central galaxy is not completely aligned with its host
halo. We assume that the probability distribution function (PDF) of
the misalignment angle $\theta$ between the major axes of central LRGs
and their host halos is a Gaussian function with a zero mean and a
width $\sigma_\theta$,
\begin{equation}
  f(\theta;\sigma_\theta)d\theta = \frac{1}{\sqrt{2\pi}\sigma_\theta}
  \exp{\left[-\frac{1}{2}\left(\frac{\theta}{\sigma_\theta}\right)^2\right]}d\theta, \label{eq:misalignment}
\end{equation}
where $\sigma_\theta$ is the misalignment angle parameter or the
typical misalignment angle.  We artificially assign misalignment to
position angles of each mock central LRG according to
equation(\ref{eq:misalignment}) before the ellipticity correlation
function is measured. It is expected that the larger the value of
$\sigma_\theta$ is, the smaller the amplitude of the ellipticity
correlation function systematically becomes. For each chosen value of
$\sigma_\theta$ and each LRG mock sample, we generate nine misaligned
LRG samples by choosing different random seeds in order to refine the
statistics of our model predictions. Although the obtained nine mock
ellipticity catalogs are not independent each other, their average can
reduce random fluctuation from different random seeds. Finally, our
model prediction for each misalignment parameter $\sigma_\theta$ is
calculated by averaging over $7\times 9=63$ misaligned samples.

In comparing the observational data with the model prediction, we first
compute the model ellipticity correlation function with a given
parameter of $\sigma_\theta$ to be tested.  $\chi^2$ statistics are
then calculated as
\begin{equation}
  \chi^2(\sigma_\theta) = \sum_{i,j} \Delta c_{11}(r_i;\sigma_\theta)
  C_{ij}^{-1} \Delta c_{11}(r_j;\sigma_\theta),
\end{equation}
where $C_{ij}$ is the covariance matrix given by equation
(\ref{eq:cov}), $\Delta c_{11}(r_i;\sigma_\theta)$ the difference
between the observed and the model values in the $i$th separation bin,
and $i$ and $j$ runs over the number of bins. In this analysis the
number of bins is 8 and the degree of freedom is 7. Thus the 99
realizations constructed from jackknife resampling are large enough to
derive a nonsingular matrix, as was already stated in Section
\ref{sec:ecf}. The range of $\sigma_\theta$ in our calculation of
$\chi^2$ is $20<\sigma_\theta<50^{\circ}$ with the width of $\Delta
\sigma_\theta =1^{\circ}$. The binned values of $\chi^2$ are then
cubic-spline interpolated.

\begin{figure}
\epsscale{1.1}
\plotone{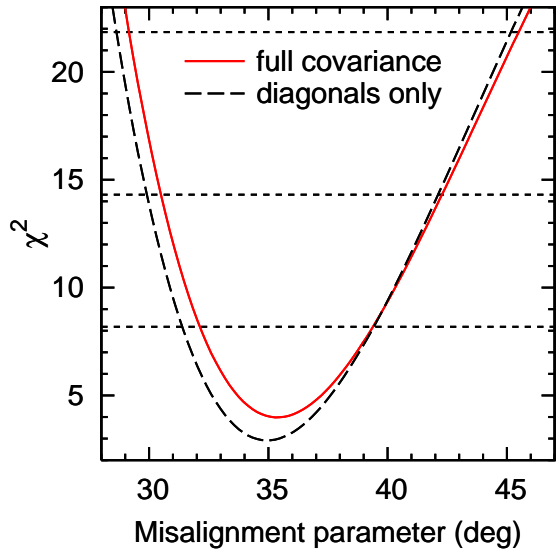}
\caption{$\chi^2$ distribution for misalignment angle parameter
$\sigma_\theta$. The best-fit parameter is
$\sigma_\theta=35.4^{+4.0}_{-3.3}$ deg ($68\%$ C.L.) when we use the
full covariance matrix, while $\sigma_\theta=35.0^{+4.4}_{-3.6}$ deg
when we use only the diagonal elements of the covariance. The minimum
value of $\chi^2$ is $\chi^2_{\rm min}=3.983$ and $2.915$ with $7$
dof, respectively.  The horizontal dotted lines show 68\%, 95\%, and
99\% confidence levels.\\~ }
\label{fig:chi2}
\end{figure}

Figure \ref{fig:chi2} shows $\chi^2$ as
a function of the misalignment parameter $\sigma_\theta$. For
comparison, both the results that all the elements and only diagonals
of the covariance matrix are used are given.  These two results are in
very good agreement, indicating that the non-diagonal elements of the
error matrix are not important.
The fits of the observed ellipticity correlation function to the model
prediction using the full covariances give $\sigma_\theta =
35.4^{+4.0}_{-3.3}$ (68\% confidence level), and $\chi^2_{\rm
min}=3.983$ with $7$ dof. 

%%%%%%%%%%%%%%%%%%%%%%%%%%%%%%%%%%%%%%%%%%%%%%%%%%%%%%%
%%%%%%%%%%%%%%%%%%%%%%%%%%%%%%%%%%%%%%%%%%%%%%%%%%%%%%%

The model prediction of $c_{11}$ with $\sigma_\theta=35^{\circ}$ is
shown in the top panel of Figure \ref{fig:vardelta}. As a cross-check,
we also plot the model of $c_{22}$ with the same $\sigma_\theta$ in
the bottom panel of Figure \ref{fig:vardelta}, which is also in very
good agreement with the observed $c_{22}$. This accordance
additionally enhances the validity of our analysis.

Recently there were two papers by \citet{Kang2007} and
\citet{Wang2008} who studied the misalignment angle between galaxies
and their host halos. Although they used the same observed statistics
of the alignment angle between the major axis of the central galaxies
and their connecting lines to satellites\citep{Yang2006}, they
obtained the typical misalignment angle with different results (about
$40^{\circ}$ by \citet{Kang2007} and $23^{\circ}$ by \citet{Wang2008}
for the whole sample of blue and red central galaxies in their
papers). The difference may come from their different methods to trace
the satellite spatial distribution in their modeling. \citet{Kang2007}
have used a semi-analytical model to trace satellites, and
\citet{Wang2008} have first tried to determine the spatial
distribution of satellites within halos. The discrepancy might come
from the fact that the triaxial shape of satellite distribution within
groups determined by \citet{Wang2008} is much rounder than dark matter
halos in simulations \citep{JS2002,Kang2007}.  Our analysis does not
need to make any assumption for the satellite galaxies or the shape of
halos. Because LRGs are red centrals, our results should be compared
with the misalignment value $16.6\pm 0.1$ degrees for red centrals in
\citet{Wang2008}(their Table 2). Their value is significantly smaller
than ours, and such a small value is strongly rejected by our
analysis.

\section{Conclusions}\label{sec:conclusion}

We have measured the ellipticity auto- and cross-correlation functions
$c_{ab}(r)$ of the spectroscopic LRG sample from the SDSS. We have
detected positive alignment between pairs of the LRGs up to $\sim 30
\himpc$ scales.  More luminous LRGs have a stronger correlation of
ellipticity than less luminous ones although the error bars are large,
while no significant evidence is found for redshift dependence between
redshifts 0.2 and 0.4.

Accurate measurement of spectroscopic redshifts enables us to divide
our LRG sample into centrals and satellites.  In order to examine the
existence of misalignment between central LRGs and their host dark
matter halos, we employed a high-resolution $N$-body
simulation. Adopting the best-fit HOD parameters of the LRGs already
obtained in previous studies, we assigned central and satellite LRGs
to simulated halos. Then we measured the model ellipticity correlation
functions from the mock central LRGs. The ellipticity correlation is
predicted to have the same shape as, but an amplitude about 4 times
higher than, our observation if the central LRGs are perfectly aligned
with their host halos.

We assumed misalignment of the central LRGs with their parent dark
halos to follow a Gaussian distribution with zero mean and a width
$\sigma_\theta$.  By comparing the observed ellipticity correlation
function $c_{11}$ with its model predictions, we have obtained a tight
constraint on the misalignment parameter as $\sigma_\theta =
35.4^{+4.0}_{-3.3}$. A model that the LRGs and host halos are
completely aligned was strongly rejected in our analysis. This is an
accurate detection of the misalignment using observed data.

The results have important implications for weak lensing
observations. To be specific for the current discussion, we assume
that the source galaxies are at redshift about 1 and have limiting
magnitude $R_{AB}=24.5$ similar to that of a recent large CFHTLS weak
lensing survey \citep{Fu2008}. From the HOD analysis by
\citet{Zheng2007} for the clustering of galaxies in the DEEP2 redshift
survey which has a similar limiting magnitude, we know that more than
$\sim 80\%$ of these source galaxies are central galaxies in dark
matter halos $\sim 4\times 10^{11}h^{-1}M_{\sun}$. If these central
galaxies have the same misalignment distribution relative to their
host halos as the LRGs, we expect that the intrinsic ellipticity
correlation of the galaxies is about 4 times smaller than that of the
hosting halos. Considering that the distribution of the intrinsic
ellipticity of the galaxies at $z\approx 1$ is quite similar to that
of host halos (G. Gao et al. 2009, in preparation), we expect that
this intrinsic ellipticity correlation of galaxies can contribute by
about 5\% to the shear correlation according to the ellipticity
correlation function of halos \citep[Figure 3 of][]{Jing2002} and the
shear power spectrum \citep[e.g.,][]{Wittman2000, Kaiser2000}. The
contribution is decreased if the blue disk galaxies are misaligned
more with their host halos than the red ones
\citep[see,][]{Yang2006}. A more precise assessment of the impact on
the weak lensing observations will be discussed in a future paper
based on more detailed modeling of galaxies in halos.

The misalignment distribution of the LRGs can also serve as a test for
the formation of giant ellipticals. The dry and gas-rich mergers may
result in different properties of elliptical galaxies, including their
orientations. This may also serve a test for the hierarchical
formation scenario of galaxies. We will compare this observation with
the central galaxies in a hydro/$N$-body cosmological simulation in a
future paper.

The current SDSS-II Legacy Survey will be completed in the next Data
Release.  However, an upcoming survey, the SDSS-III's Baryon
Oscillation Spectroscopic Survey
(BOSS)\footnote{http://www.sdss3.org/cosmology.php}, is planned to
intensively survey the LRGs at the more distant universe, $0.5<z<0.8$.
Although the main goal of the BOSS is accurate measurement of
cosmological distance scales using baryon acoustic oscillations
\citep[e.g.,][]{E2005}, it will also enable us to discuss the
evolutionary effects of LRGs. When LRGs at wider redshift ranges are
available, we will be able to improve the results obtained in this
work. Then the luminosity and redshift dependences of misalignment
between central galaxies and their host halos will be discussed more
accurately, which will lead to a better understanding of galaxy
formation and evolution.

\acknowledgments

We thank Masahiro Takada for useful discussion on weak
lensing. T.O. thanks Issha Kayo for valuable conversations about the
SDSS data, and Y.P.J. thanks Xiaohu Yang for his helpful discussion on
the misalignment results from the SDSS group catalog. This work is
supported by NSFC (10533030, 10821302, 10878001), by the Knowledge
Innovation Program of CAS (No. KJCX2-YW-T05), and by 973 Program
(No.2007CB815402).  Numerical calculations are in part performed on a
parallel computing system at Nagoya University.

Funding for the SDSS and SDSS-II has been provided by the Alfred
P. Sloan Foundation, the Participating Institutions, the National
Science Foundation, the U.S. Department of Energy, the National
Aeronautics and Space Administration, the Japanese Monbukagakusho, the
Max Planck Society, and the Higher Education Funding Council for
England. The SDSS Web Site is http://www.sdss.org/

The SDSS is managed by the Astrophysical Research Consortium for the
Participating Institutions. The Participating Institutions are the
American Museum of Natural History, Astrophysical Institute Potsdam,
University of Basel, University of Cambridge, Case Western Reserve
University, University of Chicago, Drexel University, Fermilab, the
Institute for Advanced Study, the Japan Participation Group, Johns
Hopkins University, the Joint Institute for Nuclear Astrophysics, the
Kavli Institute for Particle Astrophysics and Cosmology, the Korean
Scientist Group, the Chinese Academy of Sciences (LAMOST), Los Alamos
National Laboratory, the Max-Planck-Institute for Astronomy (MPIA),
the Max-Planck-Institute for Astrophysics (MPA), New Mexico State
University, Ohio State University, University of Pittsburgh,
University of Portsmouth, Princeton University, the United States
Naval Observatory, and the University of Washington.

\end{document}